\documentclass[aoas,MSNbibl,nameyear,seceqn,dvips]{arximspdf}

\doi{10.1214/13-AOAS697} 
\volume{8}
\issue{1}
\pubyear{2014}
\firstpage{481}
\lastpage{498}

\makeatletter

\newtheorem{thmm}{Theorem}[section]
\newtheorem{prop}[thmm]{Proposition}
\newproclaim{remark}[thmm]{Remark}
\makeatother

\begin{document}
\begin{frontmatter}

\title{Replicability analysis for genome-wide association~studies}
\runtitle{Replicability analysis for GWAS}

\begin{aug}
\author{\fnms{Ruth} \snm{Heller}\corref{}\thanksref{t1}
\ead[label=e1]{ruheller@post.tau.ac.il}}
\and
\author{\fnms{Daniel} \snm{Yekutieli}
\ead[label=e2]{yekutiel@post.tau.ac.il}}
\address{Department of Statistics and Operations Research\\
Tel-Aviv University\\
Tel-Aviv\\
Israel\\
\printead{e1}\\
\phantom{E-mail:\ }\printead*{e2}}
\thankstext{t1}{Supported by the Israel Science Foundation (ISF)
Grant no. 2012896.}
\runauthor{R. Heller and D. Yekutieli}
\affiliation{Tel-Aviv University}
\end{aug}

\received{\smonth{9} \syear{2012}}
\revised{\smonth{7} \syear{2013}}

%
\begin{abstract}
The paramount importance of replicating associations is well
recognized in the genome-wide associaton (GWA) research community,
yet methods for assessing replicability of associations are scarce.
Published GWA studies often combine separately the results of
primary studies and of the follow-up studies. Informally, reporting
the two separate meta-analyses, that of the primary studies and
follow-up studies, gives a sense of the replicability of the
results. We suggest a formal empirical Bayes approach for
discovering whether results have been replicated across studies, in
which we estimate the optimal rejection region for discovering
replicated results. We demonstrate, using realistic simulations,
that the average false discovery proportion of our method remains
small. We apply our method to six type two diabetes (T2D) GWA
studies. Out of 803 SNPs discovered to be associated with T2D using
a typical meta-analysis, we discovered 219 SNPs with replicated
associations with T2D. We recommend complementing a meta-analysis
with a replicability analysis for GWA studies.
\end{abstract}

%

\end{frontmatter}

\section{Introduction}

The aim of a genome-wide association (GWA) study is to identify
genetic variants that are associated with a given phenotype. An
analysis that combines several GWA studies of the same phenotype
may have increased power to discover the genetic variants that are
associated with the phenotype. Such a meta-analysis combines all the
data from all the studies to compute
an overall $p$-value for each SNP. The overall $p$-values are used
to identify the loci that are associated with the disease. A seminal
example of combining data to identify association comes from the
field of type 2 diabetes (T2D) GWAS. \citet{voight10} discover in a
meta-analysis single nucleotide polymorphisms (SNPs) associated with
T2D that were not discovered in single studies.

The paramount importance of replicating associations has been
well recognized in the GWAS literature
[e.g., \citet{McCarthy08,NCI07}]. \citet{kraft09} note that for
common variants, the anticipated effects are modest and very similar
in magnitude to the subtle biases that may affect genetic
association studies---most notably population stratification bias.
For this reason, they argue that it is important to see the
association in other studies conducted using a similar, but not
identical, study base. \citet{Ioannidis11} discuss multiple steps
needed to validate ``omics'' findings, including ``replication,''
which they define as the step to answer the question ``Do many different
data sets and their combination
(meta-analysis) get consistent results?''

Meta-analysis of several GWA studies aims to discover the associations
that are
present in at least one study, not replicated associations. We
define \emph{replicability analysis} as an analysis with the aim to
discover replicated associations, that is, associations between SNP and
phenotype that are present in more than one of the studies.
Meta-analysis methods are not appropriate for discovering replicated
associations. To see this, consider the scenario where for testing
the null hypothesis that a SNP is independent of the phenotype, the
$p$-value is extremely small in one study, but not small at all in
the other studies. The meta-analysis will result in a small combined
$p$-value, since there is evidence of association of this SNP with
the phenotype, but there is no evidence that this association is
replicated. Therefore, a~small $p$-value in a typical meta-analysis
is evidence toward association of the SNP with the phenotype in at
least one study, but it is not evidence that the association has
been replicated in more than one study.

Many methods exist for meta-analysis, where follow-up studies simply
serve to add power.
See \citet{Hedges85}, \citet{BY05}, \citet{skol06} and
\citet{Cardon07}, among others. However, only a handful of methods
have been suggested so far for replicability analysis.
Benjamini, Heller and Yekutieli [(\citeyear{benjamini09}); hereafter, BHY09]
suggest applying the Benjamini--Hochberg procedure
[\citet{yoav1}], henceforth referred to as the BH procedure, on
partial conjunction hypotheses $p$-values introduced in \citet{conj}.
Bogomolov and Heller (\citeyear{Bogomolov12}) focus on replicability analysis for two studies,
and suggest an alternative false discovery rate (FDR) controlling
procedure for this setting. \citet{Natarajan12} suggest a
list-intersection test to compare the top-ranked gene lists from
multiple studies in order to discover a common significant set of
genes. In this work, we suggest an empirical Bayes approach to
replicability analysis. This approach may be viewed as an extension
of the empirical Bayes approach of \citet{efron08}. We estimate the
local Bayes FDRs under the various configurations of association
status of SNP with phenotype across studies, and then sum up the
relevant probabilities in order to estimate the Bayes FDR.


The motivating example for this work came from the field of T2D GWA
studies and, therefore, we discuss this work in the context of GWA
studies. However, the proposed approach is a general approach for
assessing replicability in several studies when each study examines
the same hypotheses. Section~\ref{sec-example} describes the
motivating example and defines formally our replicability analysis
aim. In Section~\ref{sec-replicability} we present the empirical
Bayes method, and in Section~\ref{sec-example2} we apply the method
to the motivating example. In Section~\ref{sec-sim} we use simulations to
evaluate the performance of our method. We show that in realistic
simulations, the average false discovery proportion
(FDP) of our method remains small, while the power is much greater
than the power of the method of BHY09. A similar
observation was made in \citet{sun11}, where the advantage of using
an empirical Bayes approach to testing sets of hypotheses over the
method of \citet{conj} was illustrated by an application to
time-course microarray data. We conclude with a brief summary in
Section~\ref{sec-discussion}. 
\section{Motivating example and formulation of the
replicability analysis aims}\label{sec-example}
Voight et al.
(\citeyear{voight10})
conducted a meta-analysis of eight T2D GWA studies comprising 8130
T2D cases and 38,987 controls of European descent. They combined the
case-referent data from the Wellcome Trust Case Control Consortium
(WTCCC), the Diabetes Genetics Initiative (DGI), the Finland-US
Investigation of NIDDM genetics (FUSION) scans, deCode genetics
(DECODE), the Diabetes gene Discovery Group, the Cooperative Health
Research in the Region of Augsburg group, the Rotterdam study
(ERGO) and the European Special Populatin Research Network
(EUROSPAN). Based on a meta-analysis of these studies,
\citet{voight10} selected a few dozen SNPs for follow-up, and reported
the SNPs that had a small $p$-value in the follow-up study, saying
that these SNPs showed, in their words, ``strong evidence for
replication.''

We received permission to use the $p$-values for the following six
studies used for meta-analysis in \citet{voight10}: EUROSPAN,
DECODE, ERGO, DGI, FUSION and WTCCC. For these six studies, our aim
was to discover the SNPs that show strong evidence for replication
of association with T2D within a formal statistical analysis
framework. Replication of association can be defined in several
ways: with or without regard to the direction of association; with
at least $u$ out of the six studies showing association, where $u\in
\{2,\ldots,6\}$ is fixed in advance. Since direction consistency is
typically sought between the primary and follow-up studies in GWAS
[e.g., \citet{voight10}], our definition takes the directionality
into account. For the six studies, we consider a SNP as having a
replicated association if there is enough evidence to establish that
the association of SNP with the phenotype is in the same direction
in at least two studies.

In order to define the replicability aim formally, we use the
following notation. Suppose there are $n$ independent studies, and
in each study $M$ SNPs are measured. For SNP $j$ in study $i$, define
$H_{ij}$ as follows:
\[
H_{ij} = \cases{ %
1, &  $\mbox{if SNP $j$
is positively associated with the phenotype in study $i$},$
\vspace*{2pt}\cr
0, & $\mbox{if SNP $j$ is not associated with the phenotype in study $i$},$
\vspace*{2pt}\cr
-1, & $\mbox{if SNP $j$ is negatively associated with the phenotype in study $i$}.$}
\]

Let $T_{ij}$ be the test statistic of SNP $j$ in study $i$.
Following \citet{Efron10}, rather than computing the $p$-value, we
transform the test statistic into a $z$-score $Z_{ij} =
\Phi^{-1}(F_{i0}(T_{ij}))$, where $F_{i0}$ is the cumulative
distribution functions for $T_{ij}$ when $H_{ij}=0$ and
$\Phi^{-1}$ is the inverse of the standard normal cumulative
distribution function, respectively. The conditional density of
$Z_{ij}$ given $H_{ij}$ is
\[
f(z|H_{ij}) = \cases{ %
f_{i,1}(z), & \quad $\mbox{if } H_{ij}=1,$
\vspace*{2pt}\cr
f_0(z), &\quad  $\mbox{if } H_{ij}=0,$
\vspace*{2pt}\cr
f_{i,-1}(z), &\quad $\mbox{if } H_{ij}=-1,$}
\]
where $f_0(z)$ is the standard normal density.

Let $\mathcal{H} = \{\vec{h} = (h_1,\ldots,h_n)\dvtx h_i \in\{-1,0,1 \}
\}$ be the set of $3^n$ possible configurations of the vector of
association status (of SNP with phenotype) in the $n$ studies. We
are interested in examining null hypotheses for the $n$ studies that
are defined by subsets of $\mathcal{H}$ denoted by $\mathcal{H}^0$.
In particular, we shall examine the \emph{no association null
hypothesis $H^0_{\mathrm{NA}}$} that the SNP is not associated with the
phenotype in any of the studies,
\[
\mathcal{H}^0_{\mathrm{NA}} \dvtx \bigl\{ (0, 0, \ldots, 0 ) \bigr
\}
\]
as well as the \emph{no replicability null hypothesis $H^0_{\mathrm{NR}}$}
that the SNP is positively and negatively associated with the
phenotype in at most one study,
\[
\mathcal{H}^0_{\mathrm{NR}}\dvtx \Biggl\{ \vec{h} \dvtx \sum
_{i=1}^nI( h_i = -1) \le1 \cap
\sum_{i=1}^nI( h_i = 1) \le1
\Biggr\},
\]
where $I(\cdot)$ is the indicator function.

Our primary goal in this work is to discover as many SNPs as
possible with false $H^0_{\mathrm{NR}}$. This goal is
distinct from the meta-analysis goal of discovering as many SNPs as
possible with false $H^0_{\mathrm{NA}}$. For example, for $n=2$ studies,
$\mathcal{H}$ contains $3^2=9$ configurations, $\mathcal{H}^0_{\mathrm{NA}} =
\{ (0,0) \},
\mathcal{H}^0_{\mathrm{NR}} = \{(0,0), (1,0), (0,1), (-1,0),
(0,-1), (1,-1), (-1,1) \}$, and we
aim to discover as many SNPs from
the index set $\{j\dvtx \vec{H}_j\in\mathcal{H}/\mathcal{H}^0_{\mathrm{NR}} \}$,
where $\mathcal{H}/\mathcal{H}^0_{\mathrm{NR}} = \{(1,1),\break (-1,-1)\}$. Had we
defined replicability without taking directionality into account,
the null hypothesis of interest would have been $\mathcal{H}^{0} =
\{(0,0), (1,0),\break (-1,0), (0,1),(0,-1) \}$, which aims to discover as
many SNPs as possible from the index set $\{j\dvtx \vec{H}_j\in
\{(1,1),(-1,-1), (-1,1),(1,-1) \} \}$. This aim could be pursued just
as easily as the aim that follows from our definition of
replicability, with the analysis method of the next Section~\ref{sec-replicability}, but we do not examine it here.

\section{The empirical Bayes approach to replicability
analysis}\label{sec-replicability}

\subsection{The empirical Bayes approach to multiple
testing}\label{subsec-eBayes} The two group model provides a simple
Bayesian framework for multiple testing; see, for example, Chapter~2 in
\citet{Efron10}. Each SNP in study $i$ has marginal probability
$\pi_0(i)$ of not being associated with the phenotype, that is,
$\Pr(H_{i j} = 0) = \pi_0(i)$. Conditional on $H_{ij}=0$, the SNP
has a standard normal density, $f_0(z)$. Unconditionally, the
continuous marginal (mixture) density is $f_i(z)$.
For a subset $\mathcal{Z}$ of $\Re$, let
$P_0(\mathcal{Z}) = \int_{\mathcal{Z}}f_0(z)\,dz$ and
$P_i(\mathcal{Z}) = \int_{\mathcal{Z}}f_i(z)\,dz$.

Suppose we observe $z_{ij}\in\mathcal{Z}$ and wish to test
$H_{ij}=0$. A direct application of Bayes' theorem yields
\[
Fdr_i(\mathcal{Z}) = \Pr(H_{ij}=0 | z_{ij}\in
\mathcal{Z}) = \pi_0(i)P_0(\mathcal{Z})/P_i(
\mathcal{Z}).
\]
Adopting the
terminology in \citet{Efron10}, we call $Fdr_i(\mathcal{Z})$ the
\emph{Bayes FDR} for~$\mathcal{Z}$: if we report $z_{ij} \in
\mathcal{Z}$ as nonnull, that is, if we report $H_{ij}\neq0$, then
$Fdr_i(\mathcal{Z})$ is the chance that we have made a false
discovery, that is, that $H_{ij}=0$.

Theorem~1 of \citet{storey03b} shows that for the two group model for independent
test statistics, $Fdr_i(\mathcal{Z})$ is closely connected to the FDR
introduced in \citet{yoav1}. Let $\mathbf{H}_i = (H_{i1},\ldots,H_{iM}),
\mathbf{Z}_i = (Z_{i1},\ldots,Z_{iM})$,
$Q(\mathcal{Z}, \mathbf{H}_i) = \sum_{j=1}^MI(z_{ij}\in\mathcal{Z},
H_{ij}=0)/\max(R_i,1)$, where $R_i = \sum_{j=1}^MI(z_{ij}\in
\mathcal{Z})$ is the number of $z$-scores in the rejection region.
The FDR is $\operatorname{FDR}(\mathcal{Z}, \mathbf{H}_i) = E_{ \mathbf{Z}_i|
\mathbf{H}_i}Q(\mathcal{Z}, \mathbf{H}_i)$. Taking expectation over the random
$ \mathbf{H}_i$,
\[
E_{ \mathbf{H}_i} \bigl[\operatorname{FDR}(\mathcal{Z}, \mathbf{H}_i)\bigr]=
\Pr(R_i>0)E_{ \mathbf{Z}_i,
\mathbf{H}_i}[Q|R_i>0] =
\Pr(R_i>0)Fdr_i(\mathcal{Z}).
\]

If $\mathcal{Z}$ is a single point
$z_0$, then the \emph{local Bayes FDR} is
\[
fdr_i(z_0) = \Pr(H_{ij}=0|z_{ij}=z_0)
= \pi_0(i)f_0(z_0)/f_i(z_0).
\]
$Fdr_i(\mathcal{Z})$ is the conditional expectation of $fdr_i(z)$
given $z\in\mathcal{Z}$ [\citet{efron02}],
%
\begin{equation}
Fdr_i(\mathcal{Z})= E_{f_i}\bigl(fdr_i(z)|z
\in\mathcal{Z}\bigr). \label{eq-FdrAsConditionalExpectation}
\end{equation}

The Bayes false negative rate is $Fnr_i(\mathcal{Z}) =
\Pr(H_{ij}\neq0|z_{ij}\notin\mathcal{Z})$ [\citet{Efron10}]. Similar
to \citet{storey07} and \citet{sun07}, we observe that among all
possible rejection regions $\mathcal{Z}$ constrained to satisfy that
$Fdr(\mathcal{Z})\leq q$, the region with maximal probability, and
with minimal Bayes false negative rate, will be of the form
%
\begin{equation}
\label{eq-oracle.region} \mathcal{Z}_{\mathrm{OR}} = \bigl\{z\dvtx fdr_i(z)
\leq t(q) \bigr\}.
\end{equation}
The result is stated formally in the following proposition.

\begin{prop}\label{prop1}
Assume the two group model holds for the $z$-scores in study $i$.
Let $t(q)$ in expression (\ref{eq-oracle.region}) be such that
$Fdr_i(\mathcal{Z}_{\mathrm{OR}}) = q$. For any $\mathcal{Z}$ satisfying
$Fdr_i(\mathcal{Z})\leq q$,
\begin{longlist}[1.]
\item[1.]$P_i(\mathcal{Z})\leq P_i(\mathcal{Z}_{\mathrm{OR}})$.
\item[2.]$Fnr_i(\mathcal{Z}_{\mathrm{OR}})\leq Fnr_i(\mathcal{Z})$.
\end{longlist}
\end{prop}
See the proof in Section~1 of the supplementary material
[\citet{heller14}].

In the two group model, $\pi_0(i)$ and $f_i$ are needed in order to
compute the local Bayes FDR. These quantities are estimated in the R
package \emph{locfdr}, available on CRAN. Poisson regression is used
to estimate the marginal density of the $z$-scores, $\hat f_i$. The
assumption that $z$-scores that fall in the range of the central
50\% of the null distribution are null is used to estimate the
fraction of null hypotheses: $\hat{\pi}_0(i) = \frac{|\{j\dvtx z_{ij}\in
[\Phi^{-1}(0.25),\Phi^{-1}(0.75) ]\}|}{M\times0.5}$. Other estimation methods
are suggested in \citet{strimmer08},
\citet{Muralidharan10}, \citet{storey03}, \citet{yoav2} and \citet{Jin07}.

For a rejection region $\mathcal{Z}$, equation
(\ref{eq-FdrAsConditionalExpectation}) shows that
$Fdr_i(\mathcal{Z})$ may be estimated by
\[
\widehat{Fdr}_i(\mathcal{Z}) = \frac{\sum_{j\dvtx z_{ij} \in
\mathcal{Z}}\widehat{fdr}(z_{ij})}{|\{j\dvtx z_{ij}\in\mathcal{Z}\}|},
\]
where $\widehat{fdr}_i(z_{ij})=\hat{\pi}_0(i)f_0(z_{ij})/\hat
f_i(z_{ij})$ is the estimated local Bayes FDR of $z$-score $z_{ij}$,
and $|\{j\dvtx z_{ij}\in\mathcal{Z}\}|$ is the number of $z$-scores in
$\mathcal{Z}$. The estimated optimal rejection region is
$\mathcal{Z} = \{z_{ij}\dvtx \widehat{fdr_i}(z_{ij})\leq t(q) \}$, where
$t(q)$ is the largest threshold so that
$\widehat{Fdr}_i(\mathcal{Z})$ is at most $q$.


\subsection{Generalization of the two group model}
Each SNP has probability $\pi(\vec{h})$ of having association
configuration $\vec{h}$, that is, $\Pr(\vec{H}_j = \vec{h}) =
\pi(\vec{h})$. We assume the $z$-scores are independent across
studies conditional on the association status $\vec{H}_j = \vec{h}$,
so the vector of $n$ $z$-scores $\vec{z}_j = (z_{1j},\ldots,
z_{nj})$ has density $f(\vec{z}_j|\vec{h}) = \prod_{i = 1}^n
f(z_{ij}|h_i)$.
Note that $\pi_0(i)$ is equal to the sum of the
probabilities $\pi(\vec{h})$ over all $3^{n-1}$ configurations
$\vec{h}\in\mathcal{H}$ with $h_i=0$.

Suppose we observe $\vec{z}_j$ for SNP $j$ and wish to test
$\vec{H}_j\in\mathcal{H}^0$. A direct application of Bayes' theorem
yields the local Bayes FDR
%
\begin{equation}
\label{eq-replfdr} fdr_{\mathcal{H}^0}(\vec{z}_j) = \Pr\bigl(
\vec{H}_j\in \mathcal{H}^0|\vec{z}_j\bigr) =
\sum_{\vec{h} \in
\mathcal{H}^0} \pi(\vec{h}) f(\vec{z}_{j} |
\vec{h})/f(\vec{z}_j),
\end{equation}
where $f(\vec{z}_j) = \sum_{\vec{h}\in\mathcal{H}} \pi(\vec{h})
f(\vec{z}_{j}| \vec{h})$ is the mixture density. The local Bayes FDR
for SNP $j$ for null hypothesis $H_{\mathrm{NA}}^0$ and $H_{\mathrm{NR}}^0$,
respectively, is
\[
fdr_{\mathcal{H}_{\mathrm{NA}}^0} (\vec{z}_j) = \Pr\bigl( \vec{H}_j
\in\mathcal {H}_{\mathrm{NA}}^0| \vec{z}_j\bigr)\quad
\mbox{and}\quad fdr_{\mathcal{H}_{\mathrm{NR}}^0} (\vec {z}_j) = \Pr\bigl(
\vec{H}_j \in \mathcal{H}_{\mathrm{NR}}^0|
\vec{z}_j\bigr).
\]

For a subset $\mathcal{Z}$ of $\Re^n$, if we report for $\vec{z}_j
\in\mathcal{Z}$ that $\vec{H}_j\notin\mathcal{H}^0$, then the
Bayes FDR is, as in equation (\ref{eq-FdrAsConditionalExpectation}),
%
\begin{equation}
\label{def-pFDR} Fdr_{\mathcal{H}^0}(\mathcal{Z}) = \Pr\bigl(\vec{H}_j
\in \mathcal{H}^0 | \vec{z}_j \in\mathcal{Z}\bigr) =
E_f\bigl(fdr_{\mathcal{H}^0}(\vec{z_j})|
\vec{z}_j \in\mathcal{Z}\bigr).
\end{equation}

The optimal rejection
region to discover SNPs that are nonnull, that is, $\vec{H}_j\notin
\mathcal{H}^0$, follows from the same optimality argument of
Proposition~\ref{prop1}. The rejection region with maximal
probability and minimal Bayes false negative rate among all possible
rejection regions that are constrained to have a Bayes FDR of at
most level $q$ is
%
\begin{equation}
\label{eq-repOR} \mathcal{Z}_{\mathrm{OR},\mathcal{H}^0} = \bigl\{ \vec{z} \dvtx
fdr_{\mathcal{H}^0}(\vec {z})\leq t(q) \bigr\},
\end{equation}
where $t(q)$ is such that
$Fdr_{\mathcal{H}^0}(\mathcal{Z}_{\mathrm{OR},\mathcal{H}^0}) = q$. Section~2 of the
supplementary material [\citet{heller14}] shows numerical
examples that
demonstrate the different optimal rejection regions for no
replicability null hypotheses and for no association null
hypotheses, as well as the loss in power that occurs when the
rejection region is chosen suboptimally based on $p$-values.

To test whether $\vec{H}_j\in\mathcal{H}^0$ on the $n$
studies, we need to first estimate the local Bayes FDR for the
observed $z$-scores, $\{\widehat{fdr}_{\mathcal{H}^0}(\vec{z}_k)\dvtx
k=1,\ldots, M\}$. We use these estimates to estimate the Bayes FDR
(\ref{def-pFDR}) for every $z$-score $\vec z_j\ (j=1,\ldots,M)$:
%
\begin{equation}
\label{eq-qvalue} \widehat{Fdr}_{\mathcal{H}^0}(\mathcal{Z}_j) =
\frac{\sum_{k:
\vec{z}_{k} \in
\mathcal{Z}_j}\widehat{fdr}_{\mathcal{H}^0}(\vec{z}_{k})} {
|\{k\dvtx \vec{z}_{k}\in\mathcal{Z}_j\}|},
\end{equation}
where $\mathcal{Z}_j = \{\vec{z}_k\dvtx \widehat{fdr}_{\mathcal{H}_0}(\vec{z}_k) \leq
\widehat{fdr}_{\mathcal{H}_0}(\vec{z}_j), k=1,\ldots,M\}$.
Let $\hat{t}(q)$ be the largest estimated local Bayes FDR
satisfying $\widehat{Fdr}(\mathcal{Z}_j)\leq q$. Then, our estimate
of the optimal rejection region (\ref{eq-repOR}) is $\{\vec{z}_k\dvtx \widehat{fdr}_{\mathcal{H}_0}(\vec{z}_k) \leq\hat{t}(q),
k=1,\ldots,M\}$. We conclude that SNP $k$ is nonnull, that is,
$\vec{H}_k\notin\mathcal{H}_0$, if
$\widehat{fdr}_{\mathcal{H}_0}(\vec{z}_k) \leq\hat{t}(q)$ or,
equivalently, if $\widehat{Fdr}(\mathcal{Z}_k)\leq q$.

To compute $f(\vec z_j)$, it is
necessary to specify the conditional distributions for the three
states of nature for association for each SNP in each study:
$H_{ij}\in\{-1, 0, 1\}$. This is a key difference from the analysis
of single studies, where estimation of the marginal density of the
$z$-scores does not require estimation of the conditional
distributions. In Section~3 of the supplementary material [\citet
{heller14}] we
demonstrate the necessity of estimating the conditional
distributions for the states $H_{ij}=-1$ and $H_{ij}=1$ in order to
get a good estimate of $f(\vec z_j)$ at the tails, for $\vec H_j$
with dependent components.

Next, we show how to estimate $\pi(\vec{h})$ and the conditional
$z$-score densities that are necessary for estimating
the local Bayes FDR. 

\subsection{Estimating $\pi(\vec{h})$ and the conditional
$z$-score densities}\label{subsec-kgroupmodelestimation} The
likelihood for the $z$-scores for SNP $j$ is
%
\begin{equation}
L( \vec{\pi}; \vec{z}_j, f) = \Pr(\vec{z}_j | \vec{
\pi}) = \sum_{\vec{h} \in\mathcal{H}} f( \vec{z}_j |
\vec{h})\pi(\vec{h}),
\end{equation}
where $\vec{\pi} = \{ \pi(\vec{h})\dvtx \vec{h}\in\mathcal{H},
\sum_{\vec{h}\in\mathcal H} \pi(\vec{h}) =1 \}$ is the set of
$3^n-1$ probabilities of the multi-group model we want to estimate.

The full likelihood requires both the joint
distribution of $(\vec{H}_1 \cdots \vec{H}_M)$ and, for each
study $i\ (i=1,\ldots,n)$, the joint distribution of
$(Z_{i1},\ldots,Z_{iM})$ given $(H_{i1},\ldots, H_{iM})$.
Since the
joint distribution is unknown, we consider instead the composite
likelihood, which is the product of the marginal likelihoods for the
$M$ SNPs,
\[
L^{\mathrm{CL}} ( \vec{\pi}; \vec{z}, f ) = \prod_{j=1}^M
L(\vec{\pi}; \vec{z}_j, f ).
\]
Although the composite likelihood is different than the full
likelihood, in large problems with local dependency the maximum
likelihood estimates of the composite likelihood and the full
likelihood are very similar [\citet{Cox04}]. For GWAS the assumption
of local dependency seems reasonable, since the dependency across
SNPs diminishes as the distance between the SNPs increases. In
Section~\ref{sec-sim} we verified that the composite likelihood was
indeed appropriate using simulated data with GWA dependency.

Assuming that the probabilities in $\vec\pi$ were known, the
composite likelihood could be computed if the probability
distributions of ${z}_{ij}$ given $H_{i j}\in\{-1,0,1\}$, $i=1,\ldots
,n$, were known,
since $f(\vec z_{j}|\vec H_j) = \prod_{i=1}^n f_{i,
H_{ij}}(z_{ij})$. Conditional on $H_{ij}=0$, the density of
$z_{ij}$, denoted by $f_0(\cdot)$, is indeed known to be standard
normal [in Section~\ref{sec-discussion} we discuss what can be done
when $f_0(\cdot)$ is unknown]. Mixture model density estimation methods
can be used to
estimate $f_{i,1}$ and $f_{i,-1}$ [McLachlan and Peel (\citeyear{McLachlan00})]. First, the
methods discussed in Section~\ref{subsec-eBayes} can be used to
estimate the marginal density of the $z$-scores for each study,
$f_i$, and the fraction of SNPs with no association with the
phenotype, $\pi_0(i)$. Denoting the estimates by $\hat f_i$ and
$\hat{\pi}_0(i)$, the bimodal alternative density is
$\hat f_{i,A}(z) = \frac{\hat f_i(z) -\hat{\pi}_0(i)f_0(z)}{1-\hat{\pi}_0(i)}$.
Next, the expectation maximization (EM) algorithm, detailed in
Section~4 of the supplementary material [\citet{heller14}], is used to
find $\vec{\pi}$
that maximizes the composite likelihood.

\section{Replicability analysis of T2D GWA
studies}\label{sec-example2} Our first step in this analysis is to
estimate the fraction of null hypotheses for each of
the six studies, using the \emph{locfdr} package. In two of the
studies, the estimated
fraction of null hypotheses is 1. Since a stable
estimate of the conditional distribution under the alternative could
not be extracted for these two studies, we excluded them from the
empirical Bayes analysis. Studies DECODE, DGI, FUSION and WTCCC
had estimated fractions of null hypotheses of 0.89, 0.98, 0.98 and
0.96, respectively. Figure~1 of the supplementary material [\citet
{heller14}] shows the
histogram of $z$-scores, as well as the estimated conditional
densities, for each of the six studies, as outputted from the
\emph{locfdr} package.

\textit{Binning of $z$-scores}. In the \emph{locfdr} package, the
$z$-scores are binned before the densities are estimated.
Binning is practical in our application since in the estimation of
the local Bayes FDRs for several studies, estimated conditional
densities are multiplied. The accuracy of multiplied estimates may
be far less stable without binning. Therefore, we first divide the
$z$-scores $\{z_{ij}\dvtx j=1,\ldots,M \}$ into $B$ bins of equal width.
For this application, we tried both $B=50$ and $B=120$ and received
similar results. Let $x_{i,1} \cdots x_{i,B}$ be the centers of
these bins. We assign each $z$-score $z_{ij}$ into the bin that it is
in, denoted by $\tilde{z}_{ij} \in\{ 1,\ldots,B\}$. For SNP $j$,
the probability of the vector of $n$ binned $z$-scores $\tilde{z}_j
= (\tilde{z}_{1j},\ldots, \tilde{z}_{nj})$ given configuration
$\vec{H}_j$ is
$
\tilde{f}(\tilde{z}_j|\vec{H}_j) = \prod_{i=1}^n \tilde{f}_{i,
H_{ij}}(\tilde{z}_{ij})$, where
$
\tilde{f}_{i,H_{ij}}(b) = \frac{ f_{i,H_{ij}}(x_{i,b})}{ \sum_{l=1}^B
f_{i,H_{ij}}(x_{i,l})}$. For $H_{ij}=0$, $f_{0}(x_{i,b})$ is the
standard normal density at point $x_{i,b}$. For $H_{ij}\in
\{-1,1\}$,
\begin{eqnarray*}
f_{i, 1}(x_{i,b}) &= &\cases{ %
0, &\quad
$\mbox{if } x_{i,b}\leq0$,
\vspace*{2pt}\cr
\hat f_A( x_{i,b}), &\quad $\mbox{if } x_{i,b} > 0$,}
\quad\mbox{and}\\
f_{i, -1}(x_{i,b})
&=& \cases{ %
0, & \quad$\mbox{if } x_{i,b}\geq0$,
\vspace*{2pt}\cr
\hat f_A(x_{i,b}), & \quad $\mbox{if } x_{i,b} < 0$.}
\end{eqnarray*}
The EM algorithm was used
to find $\vec{\pi}$ that maximizes the composite likelihood on the
binned $z$-scores, $\prod_{j=1}^M \sum_{\vec{h} \in\mathcal{H}}
\tilde{f}( \tilde{z}_j | \vec{h})\pi(\vec{h})$.

For $n=4$ studies, the sets $\mathcal{H}$ and $\mathcal{H}_{\mathrm{NR}}^0$
contain, respectively, 81 and 21 configurations, and
$\mathcal{H}_{\mathrm{NA}}^0$ contains only the configuration $(0,0,0,0)$.
The empirical Bayes analysis at level $q=0.05$ discovered 803 SNPs
associated with T2D and 219 SNPs with replicated association with
T2D. A list of the 219 SNPs with replicated associations discovered
by the empirical Bayes analysis, sorted by positions on the
chromosome, is given in Section~5 of the supplementary material [\citet
{heller14}].
SNPs with replicated association included 16 distinct genes.
We extracted the SNP with smallest estimated local Bayes FDR among all
SNPs within each of these 16 genes, as well as among all SNPs in
noncoding areas.
In Table~\ref{tab-T2D} we list these 17 SNPs, along with
the estimated Bayes FDR for replicability analysis (column 5) and for
the analysis to discover association in (column 6). As expected, the
estimated Bayes
FDR is larger for replicability analysis than for an analysis to
discover associations, and the ranking for replicability is
different than for discovering associations. For example, the
empirical Bayes analysis for KIF11 ranks it $7$th for evidence of
replicability but $5$th for evidence of association; KCNJ11 is
ranked $5$th for evidence of replicability but $8$th for evidence of
association. The SNP which has by far the strongest evidence of
association, and replicated association, is in TCF7L2. This
association has been well established in previous papers
[\citet{voight10}]. The very small estimated Bayes FDRs for this SNP
are a result of compounding the strong evidence against the null
from four studies.

As a comparison procedure, we considered the replicability analysis
suggested in BHY09, which was to apply the BH procedure
on the $M$ no replicability null hypotheses $p$-values, computed as
suggested in \citet{conj}. We applied the analysis suggested in BHY09 on
the $n=4$ studies
with estimated fraction of null hypotheses below one, as well as on all
the $n=6$ studies available.
Briefly, the recipe for computing
$p$-values for the no replicability null hypotheses was as follows.
First, for every subset of $n-1$ studies, a meta-analysis $p$-value
was computed. Then, the $p$-value for the no replicability null
hypothesis was set to be the maximum of the $n$ meta-analysis
$p$-values. Since we considered in this work a concordant version of
replicability, where the association was considered replicated only
if it was present in at least two studies in the same direction, the
$p$-value was taken to be twice the smaller of the left- and
right-sided combined $p$-values using the method of Fisher, as
suggested in \citet{Owen09}.

The replicability analysis of BHY09 at level $q=0.05$
based on the four studies discovered 447 SNPs associated with T2D
and 83 SNPs with replicated association with T2D, and based on the
six studies discovered 466 SNPs associated with T2D and 113 SNPs
with replicated association with T2D. Table~\ref{tab-T2D} shows the
adjusted $p$-values based on all six available studies in columns
seven and eight, respectively. While the meta-analysis of
BHY09 indicates that there is evidence of association
in almost all these regions, evidence of replicated association is
inferred only for five regions.

The empirical Bayes approach provides for each SNP a measure of
belief in each possible configuration $\vec{h}$, conditional on its
vector of $z$-scores. For example, the vector of $z$-scores for SNP
$rs7903146$ in gene TCF7L2 was $\vec{z} = (-8.8, -4.5, -4.4, -7.5)$
in studies DECODE, DGI, FUSION and WTCCC, respectively.\vspace*{1pt} The
estimated posterior probability was 0.98 that the configuration was
$\vec{h} = (-1,-1,-1,-1)$, conditional on the binned $z$-score
vector. The vector of $z$-scores for SNP $rs10923931$ in gene NOTCH2
was $\vec{z} = ( -3.4, -4.9, -0.12, -2.8 )$ with estimated posterior
probability 0.92 for configuration $\vec{h} = (-1,-1,0,-1)$. Table~\ref{tab-T2Dposterior} shows the estimated posterior probability
distributions for these two SNPs.\looseness=1

%
\begin{table}
\tabcolsep=0pt
\caption{For the SNPs with strongest evidence toward replicability
in 17 distinct regions discovered by the empirical Bayes
replicability analysis: the estimated Bayes FDR for replicability
and for association (column 5--6); the adjusted $p$-values from the
analysis of BHY09 for replicability and for association
(column 7--8)}\label{tab-T2D}
\begin{tabular*}{\textwidth}{@{\extracolsep{4in minus 4in}}lccccccc@{}}
\hline
& & & &\multicolumn{2}{c}{\textbf{Empirical Bayes Fdr}} &\multicolumn
{2}{c@{}}{\textbf{BHY09 adjusted $\bolds{p}$-values}}\\[-6pt]
& & & &\multicolumn{2}{c}{\hrulefill} &\multicolumn
{2}{c@{}}{\hrulefill}\\
& \textbf{chr} & \textbf{pos} & \textbf{Gene} & \textbf{Replicability} &
\textbf{Association} & \textbf{Replicability} &
\multicolumn{1}{c@{}}{\textbf{Association}} \\
\hline
rs7903146 & 10 & 114758349 & TCF7L2 & 2.40e--11 & 4.61e--22 & 0.00e$+$00 &
0.00e$+$00 \\
rs10440833 & \phantom{0}6 & 20688121 & CDKAL1 & 1.60e--05 & 8.06e--08 & 9.06e--09 &
0.00e$+$00 \\
rs5015480 & 10 & 94465559 & noncoding & 1.10e--03 & 7.74e--05 &
8.78e--04 & 1.12e--07 \\
rs4402960 & \phantom{0}3 & 185511687 & IGF2BP2 & 3.14e--03 & 6.87e--04 & 0.0205 &
3.51e--05 \\
rs5215 & 11 & 17408630 & KCNJ11 & 8.91e--03 & 4.50e--03 & 1.00e$+$00 &
0.0236 \\
rs757110 & 11 & 17418477 & ABCC8 & 9.98e--03 & 6.16e--03 & 1.00e$+$00 &
0.0267 \\
rs4933734 & 10 & 94414567 & KIF11 & 0.0111 & 2.96e--04 & 1.00e$+$00 &
1.55e--05 \\
rs10923931 & \phantom{0}1 & 120517959 & NOTCH2 & 0.0134 & 2.70e--03 & 1.00e$+$00 &
3.45e--04 \\
rs11187033 & 10 & 94262359 & IDE & 0.0189 & 2.07e--03 & 0.0186 &
7.07e--06 \\
rs319602 & \phantom{0}5 & 134222164 & TXNDC15 & 0.0202 & 7.07e--03 & 1.00e$+$00 &
0.0364 \\
rs849134 & \phantom{0}7 & 28196222 & JAZF1 & 0.0210 & 7.80e--03 & 9.84e--01 &
1.16e--03 \\
rs6883047 & \phantom{0}5 & 134272055 & PCBD2 & 0.0235 & 8.55e--03 & 1.00e$+$00 &
0.0471 \\
rs10832778 & 11 & 17394073 & B7H6 & 0.0282 & 0.0164 & 1.00e$+$00 &
1.53e--01 \\
rs13070993 & \phantom{0}3 & 12217797 & SYN2 & 0.0370 & 0.0235 & 1.00e$+$00 & 0.0369
\\
rs10433537 & \phantom{0}3 & 12198485 & TIMP4 & 0.0360 & 0.0233 & 1.00e$+$00 &
0.0386 \\
rs10113282 & \phantom{0}8 & 96038252 & C8orf38 & 0.0387 & 0.0102 & 1.00e$+$00 &
0.0408 \\
rs1554522 & 17 & 25913172 & KSR1 & 0.0436 & 0.0145 & 1.00e$+$00 &
2.13e--01 \\
\hline
\end{tabular*}
\end{table}

\begin{table}
\caption{The estimated posterior probabilities for different
configurations $\vec{h}$, conditional on the binned $z$-score of
$\vec{z}$, for two example $z$-scores: $rs7903146$ in gene TCF7L2
(column~2) and $rs10923931$ in gene NOTCH2 (column~3)}\label{tab-T2Dposterior}
\begin{tabular*}{\textwidth}{@{\extracolsep{\fill}}lcc@{}}
\hline
$\bolds{\vec{h}}$ &$\bolds{\vec{z} = (-8.8, -4.5, -4.4, -7.5)}$ & \multicolumn{1}{c@{}}{$\bolds{\vec{z} = ( -3.4,
-4.9, -0.12, -2.8 )}$} \\
\hline
$(-1, -1, -1, -1)$ & 0.980 & 0.000 \\
$( -1, -1, 0, -1)$ & 0.012 & 0.924 \\
$( -1, -1, 0, 0 )$& 0.000 & 0.047 \\
$( -1, 0, -1, -1)$ & 0.008 & 0.000 \\
$( -1, 0, 0, -1)$ & 0.000 & 0.004 \\
$(0, -1, 0, -1)$ & 0.000 & 0.024\\
$( 0, -1, 0, 0)$ & 0.000 & 0.001 \\
\hline
\end{tabular*}
\end{table}

\section{Simulation studies}\label{sec-sim}
If all parameters were known, the optimal rejection region could be
calculated. In Section~2 of the supplementary material [\citet
{heller14}], we present two simple examples that demonstrate the
difference between the optimal rejection region for a replicability analysis
and that for an analysis to discover associations, and show that the
optimal region can be much larger than that based on $p$-values. Since
the optimal rejection region has to be estimated in practice,
we examine here the empirical Bayes approach that estimates the optimal
rejection region for inference. Specifically,
the goal of the simulations was twofold: First, to investigate the
effect of the number of SNPs $M$, and the dependence across SNPs, on
the empirical Bayes procedure; Second, to compare the empirical
Bayes procedure to the replicability analysis of BHY09
at the same level~$q$. In the empirical Bayes analysis, the
$z$-scores were first binned, using $B=50$ bins, and SNPs were
considered discovered if the estimated Bayes FDR in equation
(\ref{eq-qvalue}) was below $q=0.05$. In addition to the empirical
Bayes procedure that estimates $\vec\pi$ via the EM algorithm, we
also considered the oracle Bayes procedure that knows the
association status $H_{ij}$ of each SNP. The oracle Bayes procedure
estimates the conditional probabilities of the binned $z$-scores in
each study by the relative frequency of each bin conditional on the
association status, and uses the true vector $\vec\pi$ for
computing the local Bayes FDRs.
\subsection{Independence within each study}
We considered $n=3$ studies, with 2000 cases and 2000 referents
and $M \in\{ 10^3, 10^4, 10^5\}$ SNPs in each study.
Although there were $3^n = 27$ possible configurations of the vector
of associations status, our data generation process had positive
probability only for the 15 configurations that do not have a
positive and negative association for the same SNP: configuration
$(0,0,0)$ for 90\% of the SNPs; the six configurations with exactly
one true association, that is, $\vec{H}_j$ s.t.
$\sum_{i=1}^3|H_{ij}|=1$, each for 1\% of the SNPs; the eight
configurations with at least two true associations in the same
direction, that is, $\vec{H}_j$ s.t. $|\sum_{i=1}^3 H_{ij}|\geq2$, each
for 0.5\% of the SNPs. Following \citet{wakefield07}, we simulated
data for every SNP independently with disease risk, $p_{ij}$, given
by the logistic regression model $\operatorname{logit} (p_{ij}) =
\alpha+u\theta_{ij}$, where $u=0,0.5$, and 1 corresponds to 0,1 and
2 copies of the mutant allele, respectively. We sample $\theta_{ij}$
given $H_{ij}$ as follows:
\[
\theta_{ij}|H_{ij} \sim\cases{ %
U(0.25,0.5), & \quad$\mbox{if } H_{ij}=1,$
\vspace*{2pt}\cr
0, & \quad$\mbox{if } H_{ij}=0,$
\vspace*{2pt}\cr
U(-0.5, -0.25), & \quad$\mbox{if } H_{ij}=-1,$}
\]
where $U(a,b)$ denotes the uniform distribution between $a$ and $b$.
Moreover, the minor allele frequency (MAF) for each SNP $j$ in
study $i$ was sampled from $ U(0.05,0.50)$, and we set $\alpha=-6$,
so $e^{\alpha} = 0.0025$ was the prior odds of a disease due to a
SNP with $u=0$.

\textit{Results}. The simulation results were based on 50 repetitions
for $M= 10^5$, and on 100 repetitions for $M=10^4$
and $M=1000$. Figure~2 in the supplementary material [\citet{heller14}]
shows the FDP
in an analysis to discover associations and in a replicability
analysis. The variation in FDP decreases with $M$ and is very small
for $M=10^5$. Table~\ref{sim-tab1} presents the average FDP and
number of rejections, $R$. Although the average FDP of the empirical
Bayes analysis was below 0.05 for $M\geq10^4$, the average FDP
when $M=1000$ was 0.071, with a standard error (SE) of 0.006. The
empirical Bayes analysis makes only a few more discoveries than the
analysis of BHY09 when the aim is to discover
associations, but threefold more discoveries when the aim is to
discover replicated associations. For example, for $M=10^5$ SNPs
the empirical Bayes analysis discovers on average 2040 SNPs with
replicated associations, while the analysis of BHY09
discovers only an average of 684 SNPs. A comparison of columns 6
and 8 shows that the oracle Bayes analysis produces only a few more
discoveries than the empirical Bayes analysis, suggesting that the
loss of power in the estimation of the parameters is small.

%
\begin{table}
\tabcolsep=0pt
\caption{The average FDP and number of rejections $R$
in an empirical Bayes analysis
(columns 3 and 6), in the analysis of BHY09 (columns 4
and 7) and in an oracle Bayes analysis (columns 5 and 8), for
different values of $M={}$number of hypotheses}\label{sim-tab1}
\begin{tabular*}{\textwidth}{@{\extracolsep{4in minus 4in}}lccccccr@{}}
\hline
&& \multicolumn{3}{c}{\textbf{FDP (SE}${}\bolds{\times}{}$\textbf{1000)}} &
\multicolumn{3}{c@{}}{\textbf{R (SE)}}\\[-6pt]
&& \multicolumn{3}{c}{\hrulefill} &
\multicolumn{3}{c@{}}{\hrulefill}\\
\textbf{Analysis} & & \textbf{Empirical} & & \textbf{Oracle} &
\textbf{Empirical} & & \multicolumn{1}{c@{}}{\textbf{Oracle}} \\
\textbf{type} &$\bolds{M}$ & \textbf{Bayes} & \textbf{BHY09}&
\textbf{Bayes}& \textbf{Bayes} &\textbf{BHY09}&\multicolumn{1}{c@{}}{\textbf{Bayes}}\\
\hline
Replicability & $10^5$ & 0.049 (1) & 0.001 (0) & 0.050 (1) & 2040.6
(6.3) & \phantom{0}684.1 (3.4) & 2091.6 (4.8)\\
&$10^4$ & 0.049 (2) & 0.000 (0) & 0.049 (1) & \phantom{0}203.6 (1.4) & \phantom{000.}68 (0.9) &
211.2 (1.1)\\
&$10^3$ & 0.071 (6) & 0.000 (0)& 0.044 (4) & \phantom{00}20.5 (0.4) & \phantom{000}7.1 (0.3) &
22.7 (0.3) \\[3pt]
Association &$10^5$ & 0.046 (0) & 0.039 (0) & 0.050 (0) & 5911.3 (8.7)
& 5495.8 (7.8) & 6047.0 (9.3)\\
&$10^4$ & 0.047 (1) & 0.038 (1) & 0.050 (1) & \phantom{0}591.3 (1.7) & \phantom{0}549.7 (1.8)
& 610.6 (1.8)\\
&$10^3$ & 0.051 (2)& 0.040 (3)& 0.045 (2) & \phantom{00}58.7 (0.6)& \phantom{00}54.9 (0.6) &
66.6 (0.5) \\
\hline
\end{tabular*}
\end{table}

\begin{remark}
Table~\ref{sim-tab1} shows that the average FDP for the analysis of
BHY09 when the aim is to discover associations was
lower than $\pi(0,0,0)\times0.05 = 0.045$. For example, for
$M=10^5$ the average FDP was 0.039. This is due to the
discreteness of the distribution of the $p$-values that were
computed from contingency tables. Indeed, when the sample size was
tripled, the $p$-values from true no association null hypotheses
were closer to uniform and, therefore, the average FDP was closer to
the nominal level (not shown). However, the over-conservativeness
of the replicability analysis remained severe when the sample size
was tripled.
\end{remark}

\subsection{GWA dependency within each study}
We simulated three GWA studies from the simulator HAPGEN2
[\citet{Su11}]. The three studies where generated from three samples
of the HapMap project [\citet{HapMap03}]: a sample of 87 individuals
with African ancestry in Southwest USA (ASW), a sample of 165 Utah
residents with Northern and Western European ancestry (CEU), and a
sample of 109 Chinese in Metropolitan Denver, Colorado (CHD). We
limited ourselves to chromosomes 1--4, that contained $M=415\mbox{,}154$
SNPs. In these populations, the number of causal SNPs was 26 for
ASW, 22 for CEU and 27 for CHD. Since the effects are typically
small for GWA studies, we consider for each population four
subpopulations, and within each subpopulation about $1/4$ of the
causal SNPs had an increased multiplicative relative risk of 1.5.
Overall, there were 48 different causal SNPs in the four
chromosomes, out of which 22 SNPs were causal in more than one
population. Specifically, the three populations had five causal SNPs
in common and, in addition, the number of causal SNPs in common in
exactly two of the three populations was as follows: four for ASW and CEU,
seven for ASW and CHD, and six for CEU and CHD. Each study contained
8000 cases and 8000 referents from each population.
The simulator HAPGEN2 uses an estimate of the fine-scale
recombination rate map to simulate haplotypes conditional on the
reference haplotype data from the HapMap project. The simulator
assumes a hidden Markov model and treats the recombination rates and
mutation rates as transition and emission probabilities,
respectively. The resulting simulated data has the same linkage
disequilibrium (LD) patterns as each reference data from the HapMap
project.

Due to LD, the number of SNPs associated with the phenotype in
every study was larger than the number of causal SNPs. Since it is
not known from the data generation process which SNPs are associated
with the phenotype in each study, then for a noncausal SNP $j$ we
do not know whether $H^0 \in\{H_{\mathrm{NA}}^0$,$H_{\mathrm{NR}}^0\}$ is false,
since noncausal SNPs may have false $H^0$ due to LD patterns in the
different populations. Since a major goal in the simulations was to
assess whether the FDP is inflated, it was necessary to establish a
ground truth. We wanted to estimate a conservative ground truth that
with very high probability estimates a SNP as having a true $H^0$ if
indeed it is from $H^0$, at the possible expense of estimating a SNP
as having a true $H^0$ even if $H^0$ was false. The estimation of
the ground truth was as follows. The simulation studies were
repeated 20 times, resulting in 20 $p$-values per population for
every SNP. The 20 $p$-values were first combined with Fisher's
combining method, and the analysis of BHY09 was
applied to the combined $p$-values from the three populations, to
form for each SNP a combined $p$-value for $H^0 \in\{H_{\mathrm{NA}}^0$
,$H_{\mathrm{NR}}^0\}$ that is based on 20 studies per population. $H^0$ was
considered to be false for a SNP if the $p$-value for testing $H^0$
was below the severe Bonferroni threshold for FWER control at level
0.05. The resulting ground truth contains 2126 SNPs associated with
the phenotype, that is, with false $H_{\mathrm{NA}}^0$, and 695 SNPs with
replicated association with the phenotype, that is, with false
$H_{\mathrm{NR}}^0$. The ground truth based on 20 repetitions was very
similar to a ground truth that was established based on only 19 of
the 20 repetitions and, therefore, for an analysis of one repetition,
the resulting FDP using the ground truth based on 20 repetitions
was very similar to the FDP using the ground truth that results from
the 19 repetitions excluding the repetition being analyzed.

\textit{Results}. Table~\ref{tab-depsim} shows the analysis results
for the 20 repetitions of the three studies.
Although the average number of rejections was only slightly larger with
the empirical Bayes analysis than with the analysis of
BHY09 for testing associations, it was more than 20
times larger
when testing for replicated associations. The average FDP for
the empirical Bayes analysis was slightly above the nominal level
of 0.05, possibly because either ``ground truth'' was too conservative
(``false rejections'' are not really ``false'')
or the empirical Bayes analysis is indeed slightly anti-conservative
for the type of dependency that occurs in GWA
studies. Nevertheless, this simulation demonstrates the large gain
in using an empirical Bayes analysis over the analysis of
BHY09 for discovering replicated associations. This
large gain comes at a small risk, slightly inflated FDR.

\begin{table}
\caption{The average FDP, and number of rejections $R$,
in an empirical Bayes analysis
(columns 2 and 4) and in the analysis of BHY09
(columns 3 and 5), for the simulated data with GWA dependency within
each study}\label{tab-depsim}
\begin{tabular*}{\textwidth}{@{\extracolsep{\fill}}lcccc@{}}
\hline
& \multicolumn{2}{c}{\textbf{FDP (SE}${}\bolds{\times}{}$\textbf{1000)}} &
\multicolumn{2}{c@{}}{\textbf{R (SE)}}\\[-6pt]
& \multicolumn{2}{c}{\hrulefill} &
\multicolumn{2}{c@{}}{\hrulefill}\\
\textbf{Analysis type} & \textbf{Empirical Bayes} & \textbf{BHY09} &
\textbf{Empirical Bayes} & \textbf{BHY09} \\
\hline
Replicability & 0.065 (9)& 0.000 (0) & 154.1 (8.5)\phantom{0} & \phantom{0}6.4 (1.2) \\
Association & 0.072 (9) & 0.053 (5) & 274.9 (12.4) & 242.7 (10.4)\\
\hline
\end{tabular*}
\end{table}

\section{Summary}\label{sec-discussion}
In our analysis we assumed for each study that if the null
hypothesis was true for a SNP, the $p$-values were uniformly
distributed, that is, the $z$-score had a standard normal density.
\citet{efron08} lists several reasons why the empirical null may be
preferred over the theoretical null distribution of the $z$-scores.
The R package \emph{locfdr} fits the empirical null by truncated
maximum likelihood or by fitting a quadratic to $\log f_i$ near the
center. If in doubt about the theoretical null, the theoretical null
may be replaced with the empirical null in the empirical Bayes
analysis. In our analysis we estimated the conditional density of
$Z_{ij}$ given $H_{i j} \in\{ -1, 0, 1 \}$ in order to discover
replicated positive and negative associations. In future work we
intend to examine a more general parametrization of the
associations.


The accuracy of the empirical Bayes analysis relies on the ability
to estimate well the unknown parameters. We demonstrated in
simulations that the variability of the FDP decreased as the number
of hypotheses increased. In a simulation of realistic GWA studies we
demonstrated that the empirical Bayes analysis produced inferences
with a small FDP, despite the dependency among the $p$-values within
each study. A full Bayesian approach to the problem of GWA studies
replicability is not possible, since we do not know the true
likelihood. To estimate the probabilities of each of the
$3^n$ configurations of null and nonnull hypotheses, we used the
product of the marginal SNP likelihoods. In applications where the
exact likelihood is known, it is possible to use a full Bayesian
approach, so that the suggested framework for replicability analysis
can be extended to account for the uncertainty of the Bayes FDR
estimates.

From a comparison of an empirical Bayes analysis with the analysis
of BHY09, we see that they may give similar inferences
when the analysis is aimed at discovering associations. However, for
replicability the empirical Bayes analysis discovers many more
replicated associations than the analysis of BHY09.
In our analysis of the T2D studies, we removed the two studies with an
estimated fraction of null hypotheses of one from the empirical Bayes
analysis, since the alternative distribution could not be reliably
estimated for
these two studies using the R package \emph{locfdr}. However, these
studies are useful, as indicated by the fact that the analysis of BHY09
detected more associations using all 6 studies than using only the 4
studies with an estimated fraction of null hypotheses below one. How to
best incorporate these two studies into the empirical Bayes analysis is
an open question.





\section*{Acknowledgments}
We thank the principal investigators of the six T2D studies,
EUROSPAN, DECODE, ERGO, DGI, FUSION and WTCCC, for allowing us
to use their data. We also thank Shachar Kaufman for help with the
simulations, and Yoav Benjamini for very useful discussions.

\begin{supplement}[id=suppA]
\stitle{Supplementary material for
replicability analysis for genome-wide association studies}
\slink[doi]{10.1214/13-AOAS697SUPP} 
\sdatatype{.pdf}
\sfilename{aoas697\_supp.pdf}
\sdescription{Supplementary material includes the proof of
Proposition~\ref{prop1},
additional numerical examples that demonstrate the difference between
optimal rejection regions and the loss in power that occurs when the
rejection region is chosen suboptimally based on $p$-values,
discussion of the necessity to specify the
direction of the alternative for estimation of the local Bayes FDRs,
technical details of the EM algorithm,
the full table of results for the T2D example, the figure of empirical
$z$-scores for the T2D studies example, and an additional figure of
simulation results.}
\end{supplement}

%


\printaddresses

\end{document}